\documentclass[10pt,conference]{IEEEtran}

\usepackage{amsfonts}
\usepackage{url}
\usepackage{amsmath}
\usepackage{graphics}
\usepackage{color}

\newtheorem{theorem}{Theorem}

\newtheorem{lemma}{Lemma}

\newcommand{\ve}[1]{\mathbf{#1}}

\begin{document}

\title{A Better Good-Turing Estimator \\
  for Sequence Probabilities}

\author{
\authorblockN{Aaron B.~Wagner}
\authorblockA{School of ECE \\
Cornell University \\
{\tt wagner@ece.cornell.edu}}
\and
\authorblockN{Pramod Viswanath}
\authorblockA{ECE Department \\
University of Illinois \\
at Urbana-Champaign \\
{\tt pramodv@uiuc.edu}}
\and
\authorblockN{Sanjeev R.~Kulkarni}
\authorblockA{EE Department \\
Princeton University \\
{\tt kulkarni@princeton.edu }}
}

\maketitle

\begin{abstract}
We consider the problem of estimating the probability of an observed
string drawn i.i.d.\ from an unknown distribution. The key feature
of our study is that the length of the observed string is assumed to
be of the same order as the size of the underlying alphabet. In this
setting, many letters are unseen and the empirical distribution
tends to overestimate the probability of the observed letters. To
overcome this problem, the traditional approach to probability
estimation is to use the classical Good-Turing estimator.  We introduce a
natural scaling model and use it to show that the Good-Turing
sequence probability estimator is not consistent. We then introduce
a novel sequence probability estimator that is indeed consistent
under the natural scaling model.
\end{abstract}

\section{Introduction}
Suppose we are given a string drawn i.i.d.\ from an unknown
distribution. Our goal is to estimate the probability of the observed
string. One approach to this problem is to use
the {\em type}, or empirical distribution, of the
string as an approximation of
the true underlying distribution and then to calculate the resulting
probability of the observed string. It is well known that this estimator
assigns to each string its largest possible probability under an
i.i.d.\ distribution. For large enough
observation sizes, this estimator works well; indeed, for large $n$ and
a fixed underlying distribution, it is a {\em consistent} sequence probability
estimator.

Motivated by applications in natural language, we focus on a
nonstandard regime in which the size of the underlying alphabet is
of the same order as the length of the observed string. In this
regime, the type
of the observation is a poor representation of the
true probability distribution. Indeed, many letters with nonzero
probability will not be observed at all and the type will
obviously assign these letters zero probability.
This would not make for a consistent probability
estimator.

Since probability estimation and compression
are closely related, we can turn to the compression literature for
succor. The results in this literature are negative, however.
For instance, Orlitsky and Santhanam~\cite{Orlitsky:Large:DCC}
shows that universal compression of
i.i.d.\ strings drawn from an alphabet that grows linearly
with the observation size is impossible. As such, the compression
literature is unhelpful and even suggests that seeking a
consistent universal sequence probability estimator might be
futile.

Nevertheless, sequence probability estimation is of such importance
in applications that several heuristic approaches have been
developed. The foremost among them is based on the classical 
Good-Turing probability estimator (see Section~\ref{sec:Turing}). The
idea is to use the Good-Turing estimator instead of the type to
estimate the underlying probability distribution. The probability of
the sequence can then be calculated accordingly. Orlitsky et
al.~\cite{Orlitsky:Science} have studied the performance of a
similar scheme in the context of probability estimation for
patterns. No theoretical results regarding the performance of this
approach for sequence probability estimation are available, however.

To analyze the performance of this scheme, we introduce a natural
scaling model in which the number of observations, $n$,
and the underlying alphabet size grow at the same rate.
Further, the underlying
probabilities vary with $n$. The only restriction
we make is that no letter should be either too rare or too
frequent. That is, the probability that any given symbol occurs
somewhere in the string should be bounded away from 0 and 1
as the length of the string tends to infinity. In particular,
this condition requires that the probabilities of the letters
be $\Theta(1/n)$. We call this the~\emph{rare events regime}.
This scaling model is formally described in the next section.

Our model is similar to the one used by Klaassen and
Mnatsakanov~\cite{Klaassen:Shadow} and Khmaladze and
Chitashvili~\cite{Khmaladze:LNRE:Russian} to study related problems.
We used this model previously~\cite{Wagner:Large:ISIT06} to show
consistency of the Good-Turing estimate of the total probability of
letters that occur a given number of times in the observed string.
In the present paper, we use this model to first show that the 
Good-Turing estimator for sequence probabilities performs poorly; in
fact, a simple example illustrates that it is not consistent.
Drawing from this example, we then provide a novel sequence
probability estimator that improves upon the Good-Turing
estimator---in fact, we show that it is consistent in the context of
the natural scaling model. This is done in
Section~\ref{sec:better-turing}. Finally, we discuss the application
of our results to universal hypothesis testing problems in the rare
event regime in Section~\ref{sec:hypothesistesting}.

\section{The Rare Events Regime}

Let $\Omega_n$ be a sequence of finite alphabets.
For each $n$, let $p_n$ and $q_n$ be probability
measures on $\Omega_n$ satisfying
\begin{equation}
\label{assumption}
\frac{\check{c}}{n} \le \min(p_n(\omega),q_n(\omega))
  \le \max(p_n(\omega),q_n(\omega)) \le \frac{\hat{c}}{n}
\end{equation}
for all $\omega$ in $\Omega_n$,
where $\check{c}$ and $\hat{c}$ are fixed constants that
are independent of $n$.
Observe that this requires the cardinality of the alphabet
size to grow linearly in $n$
\begin{equation*}
\frac{n}{\hat{c}} \le |\Omega_n| \le \frac{n}{\check{c}}.
\end{equation*}
We observe two strings of length $n$. The first,
denoted by $\ve{x}$,
is a sequence of symbols drawn
i.i.d.\ from $\Omega_n$ according to $p_n$. The
second, denoted by $\ve{y}$, is a sequence of
symbols drawn i.i.d.\ from $\Omega_n$ according to $q_n$.
We assume that $\ve{x}$ and $\ve{y}$ are statistically
independent. Note that both the
alphabet and the underlying probability measures are
permitted to vary with $n$. Note also that by
assumption~(\ref{assumption}), each element of $\Omega_n$
has probability $\Theta(1/n)$ under both measures and
thus will appear $\Theta(1)$ times on average in both strings.
In fact, the probability of a given symbol appearing
a fixed number of times in either string is bounded
away from 0 and 1 as $n \rightarrow \infty$.
In other words, every letter is {\em rare}.
The number of distinct symbols in either
string will grow linearly with $n$ as a result.

Our focus shall be on the quantities $p_n(\ve{x})$ and
$p_n(\ve{y})$. An important initial observation to make
is that the distributions of these two random variables
are invariant under a  relabeling of the elements of
$\Omega_n$.
It is therefore convenient to consider the probabilities
assigned by the measures $p_n$ and $q_n$
without reference to the labeling of the symbols.
It is also convenient to normalize these probabilities
so that they are $\Theta(1)$.

Let $P_n$ denote the distribution of
\begin{equation*}
(np_n(x_n),nq_n(x_n)),
\end{equation*}
where $x_n$ is drawn according to $p_n$.
Likewise, let $Q_n$ denote the distribution of
\begin{equation*}
(np_n(y_n),nq_n(y_n)),
\end{equation*}
where $y_n$ is drawn according to $q_n$.

Note that both $P_n$ and $Q_n$
are probability measures on
$C := [\check{c},\hat{c}] \times [\check{c},\hat{c}]$.
It follows from the definitions
that $P_n$ and $Q_n$ are absolutely
continuous with respect to each other and the
Radon-Nikodym derivative is given by
\begin{equation}
\label{density}
\frac{dQ_n}{dP_n}(x,y) = \frac{y}{x}.
\end{equation}

Note that many quantities of interest involving
$p_n$ and $q_n$ can be computed using $P_n$ (or
$Q_n$). For example, the entropy of $p_n$ can
be expressed as
\begin{equation*}
- \int_C \log \frac{x}{n} \; dP_n(x,y)
\end{equation*}
and the relative entropy between $p_n$ and $q_n$
is given by
\begin{equation*}
D(p_n||q_n) = \int_C \log \frac{x}{y} \; dP_n(x,y).
\end{equation*}

We shall assume that $P_n$ converges in distribution
to a probability measure $P$ on $C$. Since $P_n$ and
$Q_n$ are related by~(\ref{density}), this implies
that $Q_n$ converges to a distribution $Q$ satisfying
\begin{equation*}
\frac{dQ}{dP}(x,y) = \frac{y}{x}.
\end{equation*}

\section{Problem Formulation}

Recall that the classical (finite-alphabet, fixed-distribution)
asymptotic equipartition property (AEP)
asserts that
\begin{equation}
\label{classicalAEP}
\lim_{n \rightarrow \infty} \frac{1}{n} \log \mu(\ve{w})
  = - H(\mu) \quad \text{a.s.},
\end{equation}
where $\ve{w}$ is an i.i.d.\ sequence drawn according to
$\mu$ and $H(\cdot)$ denotes discrete entropy. Loosely
speaking,~(\ref{classicalAEP}) says that the
 probability of the observed sequence, $\mu(\ve{w})$,
is approximately
\begin{equation*}
\exp(-n H(\mu)).
\end{equation*}
In the rare events regime, on the other hand,
one expects the probability of an
observed sequence to be approximately
\begin{equation*}
\left(\frac{h}{n}\right)^n
\end{equation*}
for some constant $h$.
Indeed, in the rare events  regime the following AEP holds true (all
proofs are contained in Section~\ref{Sec:proofs}).
\begin{theorem}
\label{theorem:AEP}
\begin{equation*}
\lim_{n \rightarrow \infty} \frac{1}{n} \sum_{i = 1}^n
   \log(np_n(x_i)) =
     \int_C \log(x) \; dP(x,y) \quad \text{a.s.}
\end{equation*}
\end{theorem}
\vspace{5pt}

Our goal is to estimate the limit in Theorem~\ref{theorem:AEP}
universally, that is, using only the observed sequence $\ve{x}$
without reference to the probability measures $p_n$. Of course,
in the classical setup, the analogous problem
of universally estimating the limit in~(\ref{classicalAEP})
is straightforward.
The distribution $\mu$ can be determined from the observed
sequence by the law of large numbers, from which the entropy
$H(\mu)$ can be calculated. In the rare events regime, on the
other hand, this approach fails and the problem is more
challenging.

We shall also study the following variation on this problem.
Consider the related quantity $p_n(\ve{y})$. That is, the sequence
is generated i.i.d.\ according to $q_n$, but we evaluate
its probability under $p_n$. This quantity
arises in detection problems, where one must
determine the likelihood of a given realization under multiple
probability distributions. As in the single-sequence setup,
it turns out that this probability converges if it is
suitably normalized.

\begin{theorem}
\label{theorem:AEP2}
\begin{equation*}
 \lim_{n \rightarrow \infty} \frac{1}{n} \sum_{i = 1}^n
   \log(np_n(y_i)) =
     \int_C \log(x) \; dQ(x,y) \quad \text{a.s.}
\end{equation*}
\end{theorem}
\vspace{5pt}

Our goal is then to estimate the limit in Theorem~\ref{theorem:AEP2}
using only the observed sequences $\ve{x}$ and $\ve{y}$. Again,
in a fixed-distribution setup, this problem is straightforward because
the two distributions can be determined exactly from the
observed sequences in the limit as $n$ tends to infinity. In
the rare events regime, however, the problem is more challenging.

\section{The Good-Turing Estimator}
\label{sec:Turing} The Good-Turing estimator can be viewed as an
estimator for the probabilities of the individual symbols. Let $A_k$
be the set of symbols that appear $k$ times in the sequence
$\ve{x}$, and let $\varphi_k = |A_k|$ denote the number of such
symbols. The basic form of the Good-Turing estimator assigns
probability
\begin{equation}
\label{GT}
\frac{(k+1)\varphi_{k+1}}{n \varphi_k}
\end{equation}
to each symbol that appears $k \le n - 1$ times~\cite{Good:Turing}.
The case $k = n$ must be handled separately, but this
case is unimportant to us because in the rare events regime the
chance that only one symbol appears in $\ve{x}$ is asymptotically
negligible.

The Good-Turing formula can also be viewed as an estimator for the
total probability of all symbols that appear $k$ times in $\ve{x}$,
i.e., $p_n(A_k)$. In particular, the $\varphi_k$ in the denominator
can be viewed as simply dividing the total probability
\begin{equation*}
\frac{(k+1) \varphi_{k+1}}{n}
\end{equation*}
equally among the $\varphi_k$ symbols that appear $k$ times. In
previous work, we showed that the Good-Turing total probability
estimator is strongly consistent in that for any $k \ge 0$,
\begin{align}
\nonumber
\lim_{n \rightarrow \infty} \frac{(k+1)\varphi_{k+1}}{n}
  & = \lim_{n \rightarrow \infty} p_n(A_k) \\
\label{oldresult}
  & = \int_C \frac{x^k e^{-x}}{k!} \; dP(x,y) =: \lambda_k
 \quad \text{a.s.}
\end{align}
(see~\cite{Wagner:Large:ISIT06}, where the notation is slightly different,
 for a proof of a stronger version of this
statement). The Good-Turing probability estimator in~(\ref{GT})
gives rise to a natural estimator for the probability of the
observed sequence $\ve{x}$
\begin{equation*}
\prod_{k = 1}^{n-1} \left(\frac{(k+1) \varphi_{k+1}}{n \varphi_k}\right)^
  {k \varphi_k}.
\end{equation*}
This in turn suggests the following
estimator for the limit in Theorem~\ref{theorem:AEP}
\begin{equation}
\label{turing:sequence}
\sum_{k = 1}^{n-1} \frac{k \varphi_k}{n} \log \left( \frac{(k+1)\varphi_{k+1}}
   {\varphi_k}\right).
\end{equation}
This estimator is problematic, however, because for the largest $k$
for which $\varphi_k > 0$,
\begin{equation*}
\frac{(k+1) \varphi_{k+1}}{\varphi_k} = 0,
\end{equation*}
which means that the $k$th term in~(\ref{turing:sequence})
equals $-\infty$. Various
``smoothing'' techniques have been introduced to address
related problems with the estimator~\cite{Good:Turing}.
Our approach will be to truncate
the summation at a large but fixed threshold, $K$
\begin{equation*}
\sum_{k = 1}^K \frac{k \varphi_k}{n} \log \left( \frac{(k+1)\varphi_{k+1}}
   {\varphi_k}\right).
\end{equation*}
In the rare events regime, with probability one
it will eventually happen that $\varphi_k > 0$ for all $k = 1,\ldots,K$,
thus obviating the problem. By the result in~(\ref{oldresult}),
this estimator will converge to
\begin{equation}
\label{turing:sequence:limit}
\sum_{k = 1}^K \lambda_{k - 1} \log \frac{k \lambda_k}{\lambda_{k-1}}.
\end{equation}
We next show that this quantity need not tend to the limit
in Theorem~\ref{theorem:AEP} as $K$ tends to infinity.

Let $\Omega_n$ be the set
$\{1,2,\ldots, 3n\}$.
Suppose that $p_n$ assigns probability $1/(4n)$ to the
first $2n$ elements and probability $1/(2n)$
to the remaining $n$. The distribution $q_n$
is obviously not relevant here so we shall simply set it equal to
$p_n$.

The resulting distribution $P$ will place mass $1/2$ on
each of the points $(1/4,1/4)$ and $(1/2,1/2)$.
From Theorem~\ref{theorem:AEP}, the limiting normalized
probability of $\ve{x}$ is $-(1/2) \log 8$.
By~(\ref{turing:sequence:limit}), the Good-Turing estimate converges
to
\begin{align*}
& \frac{1}{2} \sum_{k = 1}^K \frac{e^{-1/4} (1/4)^{k-1}}{(k-1)!}
   \left(1 + e^{-1/4} 2^{k-1}\right) \cdot \\
&  \phantom{\frac{1}{2} \sum}  \log\left(\frac{\sqrt{8}(1 + e^{-1/4} 2^k)}
    {4(1 + e^{-1/4} 2^{k-1})}\right)  \\
& \phantom{\frac{1}{2}} +
    \frac{1}{2} \sum_{k = 1}^K \frac{e^{-1/4} (1/4)^{k-1}}{(k-1)!}
   \left(1 + e^{-1/4} 2^{k-1}\right) \log \frac{1}{\sqrt{8}}.
\end{align*}
Now as $K$ tends to infinity, the second sum converges to 
the correct answer, $-(1/2) \log 8$. But one can verify 
that every term in the first sum is strictly positive. Thus
the Good-Turing estimator is not consistent in this example.

The problem is that the Good-Turing estimator is estimating the sum,
or equivalently the arithmetic mean, of the probabilities of the
symbols appearing $k$ times in $\ve{x}$. Estimating the sequence
probability, on the other hand, amounts to estimating the geometric
mean of these probabilities. If $p_n$ assigns the same probability
to every symbol, then the arithmetic and geometric means coincide,
and one can show that the Good-Turing sequence probability estimator
is asymptotically correct. In the above example, however, $p_n$ is
not uniform, and the Good-Turing formula converges to the wrong
value. In the next section, we describe an estimator that targets
the geometric mean of the probabilities instead of the arithmetic
mean, and thereby correctly estimates the sequence probability.

\section{A Better Good-Turing Estimator}
\label{sec:better-turing}

Write
\begin{equation*}
\overline{c} = \frac{\hat{c} + \check{c}}{2},
\end{equation*}
and then let
\begin{multline*}
\gamma_k^M = - \sum_{m = 1}^M \sum_{\ell = 0}^m
  (-\overline{c})^{-\ell} {m \choose \ell}
     \frac{(k+\ell)!}{m \cdot k!}
   \frac{(k+\ell + 1) \varphi_{k + \ell + 1}}{n} \\
  \mbox{ }
   + \log(\overline{c})
      \frac{(k+1)\varphi_{k+1}}{n}.
\end{multline*}

Note that $\gamma_k^M$ is only a function of $\ve{x}$
and in particular, it does not depend on $p_n$.
The next theorem shows that for large $K$ and $M$,
\begin{equation*}
\sum_{k = 0}^K \gamma_k^M
\end{equation*}
is a consistent estimator for the
limit in Theorem~\ref{theorem:AEP}.
\begin{theorem}
\label{theorem:one}
For any $\epsilon > 0$,
\begin{equation}
\label{consistent:one}
\lim_{n \rightarrow \infty} \left| \frac{1}{n}
  \sum_{i = 1}^n \log(np_n(x_i))
   - \sum_{k = 0}^K \gamma_k^M \right| \le \epsilon \quad \text{a.s.}
\end{equation}
provided
\begin{equation*}
\max\left(\frac{\exp(\hat{c}) \overline{c}}{\check{c}}
   \left(\frac{\hat{c} - \check{c}}
   {\hat{c} + \check{c}}\right)^{M+1}, \frac{\hat{c}^{K+1}
   c}{(K+1)!} \right) \le \frac{\epsilon}{2},
\end{equation*}
where
\begin{equation*}
c = \max(|\log \check{c}|,|\log \hat{c}|).
\end{equation*}
\end{theorem}
\vspace{5pt}

The idea behind Theorem~\ref{theorem:one} is this. Recall
from~(\ref{oldresult}) that
\begin{equation*}
\lim_{n \rightarrow \infty} \frac{(k+1) \varphi_{k+1}}{n}
  = \int_C \frac{x^k e^{-x}}{k!} \; dP(x,y) \quad \text{a.s.}
\end{equation*}
If one could find a sequence of
constants $a_k$ such that
\begin{equation*}
\sum_{k = 0}^\infty a_k \frac{x^k e^{-x}}{k!} = \log(x)
\end{equation*}
on $[\check{c},\hat{c}]$, then one might expect that
\begin{equation*}
\lim_{n \rightarrow \infty} \sum_{k = 0}^{n-1} a_k
  \frac{(k+1) \varphi_{k+1}}{n} = \int_C \log(x) \; dP(x,y) \quad \text{a.s.}
\end{equation*}
This is indeed the approach we took to find the formula for $\gamma_k^M$.

The estimator can be naturally extended to the
two-sequence setup, namely to
the problem of universally estimating $p_n(\ve{y})$.

Let $\varphi_{k,\ell}$ be the number of symbols in
$\Omega_n$ that appear $k$ times in $\ve{x}$ and
$\ell$ times in $\ve{y}$. Then let
\begin{multline*}
\tilde{\gamma}_k^M =
   - \sum_{m = 1}^M \sum_{\ell = 0}^m
     (- \overline{c})^{-\ell}
    {m \choose \ell} \frac{(k+\ell)!}{m \cdot k!}
  \sum_{j = 1}^n \frac{j\varphi_{k+\ell,j}}{n} \\
  \mbox{ } +
   \log (\overline{c})
  \sum_{j = 1}^n \frac{j \varphi_{k,j}}{n}.
\end{multline*}
Note that $\tilde{\gamma}_k^M$ is a function of $\ve{x}$ and
$\ve{y}$.

\begin{theorem}
\label{theorem:two}
For any $\epsilon > 0$,
\begin{equation*}
\lim_{n \rightarrow \infty} \left| \frac{1}{n} \sum_{i = 1}^n
   \log(np_n(y_i)) -
     \sum_{k = 0}^K \tilde{\gamma}_k^M \right| \le \epsilon \quad \text{a.s.}
\end{equation*}
provided
\begin{equation*}
\max\left(\frac{\exp(\hat{c}) \overline{c}}{\check{c}}
   \left(\frac{\hat{c} - \check{c}}
   {\hat{c} + \check{c}}\right)^{M+1},
     \frac{\hat{c}^{K+1} c}{(K+1)!}\right) \le
       \frac{\epsilon}{2}.
\end{equation*}
\end{theorem}
\vspace{5pt}

This result shows that although we are unable to  determine $p_n$
from $\ve{x}$, we are able to glean enough information about
$p_n$ to determine the limit in Theorem~\ref{theorem:AEP2}.

\section{Universal Hypothesis Testing}
\label{sec:hypothesistesting}
The $\tilde{\gamma}_k^M$ estimator leads to a natural scheme
for the problem of universal hypothesis testing. Suppose that we
again observe the sequences $\ve{x}$ and $\ve{y}$, which we
now view as training data. In addition, we observe a test
sequence, say $\ve{z}$, which is generated i.i.d.\ from the
distribution $r_n$. We assume that either $r_n = p_n$ for
all $n$ or $r_n = q_n$ for all $n$. The problem is to
determine which of these two possibilities is in effect
using only the sequences $\ve{x}$, $\ve{y}$, and $\ve{z}$.

Using Theorem~\ref{theorem:two}, one can estimate $p_n(\ve{z})$ and
$q_n(\ve{z})$, and by comparing the two, determine which of
the two distributions generated $\ve{z}$. This will make for a
 consistent universal classifier, without recourse to actually
 estimating the true underlying distributions $p_n$ and $q_n$.
 As a scheme for
universal hypothesis testing, however, this approach is quite
complicated and there is no reason to believe it would
be optimal in an error-exponent sense. We are currently
investigating other, more direct approaches to the
universal hypothesis testing problem in the rare events
regime. For a discussion of universal hypothesis testing in
the traditional, fixed-distribution regime, see
Gutman~\cite{Gutman:Classification} and
Ziv~\cite{Ziv:Classification}.

\section{Proofs}
\label{Sec:proofs}

Due to space limitations, we will only prove Theorem~\ref{theorem:AEP}
and sketch the proof of Theorem~\ref{theorem:one}. The
proofs of Theorems~\ref{theorem:AEP2} and~\ref{theorem:two}
are similar.
\begin{lemma}
\label{mean:prob}
\begin{equation*}
\lim_{n \rightarrow \infty} E\left[ \frac{1}{n}
  \sum_{i = 1}^{n} \log(np_n(x_i))
    \right] = \int_C \log(x) \; dP(x,y).
\end{equation*}
\end{lemma}
\vspace{5pt}

\begin{proof}
Note that for any $i$
\begin{align*}
E[\log(np_n(x_i))] & = \sum_{\omega \in \Omega_n} p_n(\omega)
     \log(np_n(\omega)) \\
  & = \int_C \log(x) \; dP_n(x,y).
\end{align*}
Since $\log(x)$ is bounded and continuous over $C$ and $P_n$ converges in
distribution to $P$, the result follows.
\end{proof}

\begin{lemma}
\label{conc:prob}
\begin{multline*}
\lim_{n \rightarrow \infty} \Bigg| \frac{1}{n}
   \sum_{i = 1}^n \log(n p_n(x_i)) \\
  - E\left[ \frac{1}{n} \sum_{i = 1}^{n}
    \log(np_n(x_i))\right]\Bigg| = 0 \quad \text{a.s.}
\end{multline*}
\end{lemma}
\vspace{5pt}

\begin{proof}
Consider the sum
\begin{equation*}
\sum_{i = 1}^{n} \log(n p_n(x_i)).
\end{equation*}
If one symbol in the sequence $\ve{x}$ is altered,
then this sum can change by at most
\begin{equation*}
\log \frac{\hat{c}}{\check{c}}.
\end{equation*}
It follows from the Azuma-Hoeffding-Bennett concentration
inequality~\cite[Corollary~2.4.14]{Dembo:LD} that
\begin{multline*}
\Pr\Bigg(\Bigg| \frac{1}{n} \sum_{i = 1}^{n} \log(n p_n(x_i)) \\
  - E\left[\frac{1}{n} \sum_{i = 1}^n
   \log (np_n(x_i)) \right]\Bigg| >
    \epsilon\Bigg) \\
 \phantom{\Pr\Bigg(\Bigg|\frac{1}{n} \sum}
      \le 2\exp\left(- \frac{\epsilon^2 n}{2
   (\log(\hat{c}/\check{c}))^2}\right).
\end{multline*}
The result then follows by the Borel-Cantelli lemma.
\end{proof}
Note that
Theorem~\ref{theorem:AEP} follows immediately from Lemmas~\ref{mean:prob}
and~\ref{conc:prob}.

The key step in the proof of Theorem~\ref{theorem:one} is showing
that $\gamma_k^M$ converges to the proper limit. This
is shown in the next and final lemma.
\begin{lemma}
\label{gamma:limit}
For any $\epsilon > 0$ and any $k \ge 0$,
\begin{equation*}
\lim_{n \rightarrow \infty} \left| \gamma_k^M -
  \int_C \log(x) \frac{\exp(-x) x^k}{k!} \; dP(x,y) \right|
    \le \frac{\epsilon \hat{c}^k}{k!} \quad \text{a.s.,}
\end{equation*}
provided
\begin{equation*}
\frac{\overline{c}}{\check{c}}\left(\frac{\hat{c} - \check{c}}{\hat{c} +
   \check{c}}\right)^{M+1} \le \epsilon.
\end{equation*}
\end{lemma}
\vspace{5pt}

\emph{Proof (sketch):}
Note that the limit exists by~(\ref{oldresult}).
By the triangle inequality,
\begin{multline}
\label{div:expansion}
\left|\gamma_k^M -
  \int_C \log(x) \frac{\exp(-x) x^k}{k!} \; dP(x,y)\right|
  \le \left| \gamma_k^M - \overline{\gamma}_k^M \right| \\
  \mbox{ } + \left| \overline{\gamma}_k^M  - \int_C \log(x)
      \frac{\exp(-x) x^k}{k!} \; dP(x,y)\right|,
\end{multline}
where
\begin{equation}
\label{limitgamma}
\overline{\gamma}_k^M = - \sum_{m = 1}^M \sum_{\ell = 0}^m
  (- \overline{c})^{-\ell} {m \choose \ell} \frac{(k+\ell)!}{m \cdot k!}
  \lambda_{k + \ell} + \log(\overline{c}) \lambda_k.
\end{equation}
The first term on the right-hand side of~(\ref{div:expansion})
tends to zero by~(\ref{oldresult}). Now
\begin{align*}
& - \sum_{m = 1}^M \sum_{\ell = 0}^m
  (-\overline{c})^{-\ell} {m \choose \ell} \frac{(k+\ell)!}{m \cdot k!}
    \lambda_{k + \ell} \\
  & = - \int_C \sum_{m = 1}^M
     \frac{\exp(-x) x^k}{m \cdot k!}
           (- \overline{c})^{-m}  \\
   & \phantom{= - \int_C} \cdot \sum_{\ell = 0}^m {m \choose \ell}
      \left(- \overline{c}\right)^{m - \ell}
       x^\ell \; dP(x,y).
\end{align*}
By the Binomial Theorem,
\begin{equation*}
  \sum_{\ell = 0}^m {m \choose \ell}
 \left(- \overline{c}\right)^{m-\ell} x^\ell
    = (x - \overline{c})^m.
\end{equation*}
Substituting these last two equations into~(\ref{limitgamma}) yields
\begin{multline*}
\overline{\gamma}_k^M = - \int_C \sum_{m = 1}^M
    \left(1 -  \frac{x}{\overline{c}}\right)^m
      \frac{\exp(-x) x^k}{m \cdot k!} \; dP(x,y) \\
  \mbox{ } +
  \log(\overline{c}) \lambda_k.
\end{multline*}
Using the well-known power series
\begin{equation*}
\log(1+x) = \sum_{m = 1}^\infty \frac{(-1)^{m+1}}{m} x^m,
\end{equation*}
valid for $-1 < x \le 1$, one can show that
\begin{equation*}
\sup_{\check{c} \le x \le \hat{c}} \left| \log \frac{x}{\overline{c}}
  + \sum_{m = 1}^M \frac{1}{m} \left(1 - \frac{x}{\overline{c}}\right)^m\right|
  \le \frac{\overline{c}}{\check{c}} \left( \frac{\hat{c} - \check{c}}
   {\hat{c} + \check{c}}\right)^{M+1} \le \epsilon
\end{equation*}
by hypothesis. Thus
\begin{align*}
 & \left| \overline{\gamma}_k^M - \int_C \log(x)
    \frac{\exp(-x) x^k}{k!} \; dP(x,y) \right| \\
 & = \Bigg| \int_C
    \sum_{m = 1}^M
    \left(1 - \frac{x}{\overline{c}}\right)^m
    \frac{\exp(-x)x^k}{m \cdot k!} \; dP(x,y) \\
 & \phantom{\Bigg| \int_C} + \int_C
        \log\left(\frac{x}{\overline{c}}\right)
    \frac{\exp(-x)
     x^k}{k!} \; dP(x,y) \Bigg| \\
 & \le \int_C \epsilon \ \frac{\exp(-x) x^k}{k!} \; dP(x,y)
   \le \frac{\epsilon \hat{c}^k}{k!}.
\end{align*}
\hfill \QED

Since
\begin{equation*}
\sum_{k = 0}^\infty \frac{\exp(-x) x^k}{k!} = 1,
\end{equation*}
one would expect from Lemma 3 that for large $K$ and $M$,
\begin{equation*}
\sum_{k = 0}^K \gamma_k^M
\end{equation*}
would be close to
\begin{equation*}
\int_C \log(x) \; dP(x,y).
\end{equation*}
Indeed, one can prove Theorem~\ref{theorem:one} using this approach.
The details are omitted.

\end{document}